\documentstyle[aps]{revtex}
\begin{document}
\baselineskip=1 cm
\begin{center}
{\huge {\bf Test Experiment for Time-Reversal Symmetry Breaking
Superconductivity}}
\end{center}
\vskip 0.5 cm
\baselineskip=0.55 cm
\begin{center}
{\large {\it Manfred Sigrist and Yong Baek Kim}}
\end{center}
\vskip 0.5 cm
\begin{center}
Department of Physics, Massachusetts Institute of Technology, Cambridge,
MA 02139, U.S.A.
\end{center}
\vskip 0.8 cm
\noindent
{\it A new experiment is proposed to probe the time-reversal symmetry of a
superconductor. It is shown that a time-reversal symmetry
breaking superconductor can
be identified by the observation of a fractional
flux in connection with a Josephson junction in a special geometry.}
\vskip 1 cm
\noindent
During the last few years one of the important issues in the field
of high temperature superconductivity was the experimental identification
of the symmetry of the superconducting order parameter. Various theories
suggest unconventional superconductivity due to Cooper pairing in the
``d-wave''
channel with a pair wave function of the form $ \psi({\bf k}) =
\cos k_x - \cos k_y $. The phase characteristics of this state  (the
sign change under exchange of $ x $- and $ y $-coordinate) is the
basis for a series of experiments probing directly the intrinsic
phase difference of $ \pi $ between different momentum directions
in the pair wave function \cite{1}.
It was shown that the presence of a d-wave
pairing state could also be responsible for the occurrence of the peculiar
paramagnetic response in granular
$ {\rm Bi}_2 {\rm Sr}_2 {\rm CaCu}_2 {\rm O}_8 $ (Wohlleben effect) \cite{4}.
Although by now a considerable number of experiments demonstrate convincingly
the d-wave symmetry of the superconducting state at least for
$ {\rm YBa}_2 {\rm Cu}_3 {\rm O}_7 $ (YBCO) \cite{1}, this issue is
still controversial as other experiments seem to be incompatible with
a d-wave state \cite{2}.
Recently, Kirtley et al. reported the observation
of vortices on grain boundaries in a film of (YBCO), which
carry a flux smaller than a standard flux quantum
$ \Phi_0 $ ($ = hc/2e $) (even smaller than
$ \Phi_0 /2 $ as it could originate from the intrinsic $ \pi $-phase shift
of a d-wave state) \cite{5}. Subsequently,
this observation was interpreted as an indication that the
superconducting state breaks the time-reversal symmetry $ {\cal T} $
if not in the bulk then at least near the grain boundaries \cite{6}.
Besides the high temperature superconductors,
the heavy fermion systems $ {\rm UPt}_3 $ and $ {\rm U}_{1-x}
{\rm Th}_x {\rm Be}_{13} $ ($ 0.018 < x < 0.045 $) are good
candidates to have a $ {\cal T} $-violating superconducting phase \cite{12}.
In this letter we propose a new experiment to probe the time-reversal
symmetry of a superconducting state.
In the following we will examine a concrete model which may
directly apply to the high temperature superconductors.
However, most of our conclusions are also valid for other superconductors.
We consider
a thin film of the superconducting material, which is c-axis
oriented and untwinned if any orthorhombic distortion is present
as in YBCO. (Twinning leads to various complications which we will
discuss elsewhere.)
We cut two small separate holes in the film,
which are connected by a straight homogeneous
Josephson junction (see Fig.1).
We assume that the junction is sufficiently
strong such that the Josephson penetration depth $ \lambda_J $
is much smaller than
the distance between the two holes.
Thus, if magnetic field threads the two holes then the magnetic
flux of each hole is separately a well-defined quantity.
Obviously the geometry of the
arrangement requires that the total flux of both holes together is an integer
multiple of $ \Phi_0 $. However, we claim that it is possible
to find an arbitrary amount of
magnetic flux in each hole if certain conditions
are satisfied.
Let us assume that the flux through one of the two holes is $ \Phi $.
If we applied the time-reversal operation to the system then this flux
would change sign, $ -\Phi $. If the superconducting state
(apart from the currents flowing to sustain the flux in the hole)
is invariant under time-reversal, the original and the inverted flux
should differ by an integer multiple of $ \Phi_0 $. Thus, $ \Phi
= - \Phi + n \Phi_0 $ which leads to $ \Phi = n \Phi_0 /2 $. Hence $ \Phi $
can only be an
integer (for even $ n $) or half-integer (for odd $ n $) multiple
of $ \Phi_0 $. The latter is possible only under special conditions
which are not included in our discussion \cite{4,8}. Thus, in order
to obtain a flux different from these two cases, the superconducting
state must break the time-reversal symmetry.
Next, we reflect the system at a mirror plane which includes the junction
and is parallel to the c-axis. This operation also changes the
sign of $ \Phi $. Hence, the same argument as before applies.
Both the {time-reversal} and the {\it reflection symmetry} have to
be violated in order to find an arbitrary flux $ \Phi $ in the hole.
It is reasonable
to assume that the violation of reflection symmetry
is connected with asymmetric properties
of the junction and its vicinity.
On the other hand, $ {\cal T} $ may be violated in the whole superconductor
or, alternatively, only locally near the junction as proposed by
Sigrist, Bailey and Laughlin (see also Ref.8) \cite{6}.
To be concrete let us now look at the example of a superconductor with a
two-component order parameter, $ \vec{\eta}= ( \eta_1 , \eta_2) $.
The first component
has the above introduced d-wave symmetry,
while the second is completely symmetric
(s-wave). Under tetragonal crystal symmetry $ \eta_1 $ and $ \eta_2 $ are
generally not degenerate and
have different (bare) transition temperatures
$ T_{c1} $ and $ T_{c2} $, respectively.
For our purpose it is sufficient to study this system on a
phenomenological level.
The Ginzburg-Landau free energy has the following
general form $ F = F_1 + F_2 +F_{12} $ with
\begin{equation}
F_i = \int  d^2 r [A_i(T) |\eta_i|^2 + \beta_i |\eta_i|^4 + K_i
|{\bf D} \eta_i|^2]
\end{equation}
\noindent
and
\begin{equation} \begin{array}{ll} \displaystyle
F_{12} = \int d^2 r [  & \gamma |\eta_1|^2 |\eta_2|^2 + \delta
\left( \eta^{*2}_1 \eta^2_2 + h.c. \right)  \\
& \\ & \displaystyle + K' \sum_{\alpha=x,y} s_{\alpha} \left(
(D_{\alpha} \eta_1)^* (D_{\alpha} \eta_2 )  + h.c. \right) ] ,
\end{array} \end{equation}
\noindent
where we neglect the degree of freedom related to the
$ z $-coordinate, $ s_x = +1 , s_y=-1 $, and $ A_i(T)
=a(T - T_{ci}) $ $ (i=1,2) $. The coefficients $ \beta_i $, $ \gamma $,
$ \delta $, $ K_i $ and $ K' $ are phenomenological parameters.
The symbol $ {\bf D} = \nabla - 2 \pi {\bf A}/ \Phi_0 $ denotes the
gauge invariant derivative where $ {\bf A} $ is
the vector potential. In case of $ T_{c1} >
T_{c2} $, we find $ |\eta_1|^2 (T) = - A_1(T)/2 \beta_1 $ and $ \eta_2=0 $
for $ T_{c1} > T > T^* $, while both components are
finite for $ T < T^* $. The
temperature $ T^* $ represents the renormalized transition temperature
for $ \eta_2 $, which is defined by
$ A_1(T^*) (\gamma + 2 \delta \cos (2 \theta)) = 2 \beta_1 A_2(T^*) $.
The relative phase $ \theta = \phi_1 - \phi_2 $ is determined by the
sign of $ \delta $ (we parametrize $ \eta_j = u_j e^{i \phi_j} $).
If $ \delta > 0 $ the (two-fold degenerate)
low temperature state has $ \theta = \pm \pi/2 $ and
violates $ {\cal T} $, i.e., it is the so-called $ s + id $-state.
If the superconductor has a two-component order parameter,
the property of a Josephson junction is
described by four coupling terms leading to the local current-phase relation
\begin{equation}
J= \sum_{i,j=1,2} J_{ij} \sin (\phi_{ib} - \phi_{ja})
\end{equation}
\noindent
with subscripts $ a $ and $ b $ indicating the two sides of the junction
(Fig.1).
On the other hand, the current density in the bulk
superconductor with $ \theta = \pi/2 $
is given by
\begin{equation}
J_{\alpha} = \frac{4 \pi c}{\Phi_0} \left[ R (\frac{\partial
\varphi}{\partial \alpha} - \frac{2 \pi}{\Phi_0} A_{\alpha})
+ K' s_{\alpha} \sum_{i,j=1,2} \varepsilon_{ij} u_i
\frac{\partial u_j}{\partial \alpha} \right],
\end{equation}
\noindent
where $ \varphi= \phi_1 = \phi_2 + \pi/2 $,
$ R = \sum^2_{i=1} K_i u^2_i $, and $ \varepsilon_{ij} = - \varepsilon_{ji} $
($ \alpha=x,y $).
The last term gives a finite contribution only if the
magnitude of the order parameter is varying in space.
This is the case, in general, in the vicinity of the junction,
i.e., $ u_j = u_j(\tilde{x}= {\bf r} \cdot {\bf n}) $, where $ {\bf n} $
is the normal vector of the junction (we choose $ \tilde{x}
=0 $ on the junction).
We calculate now
the flux in one of the two holes by encircling it on a path $ {\cal C} $
starting at the junction on side $ a $ and ending just
on the other side $ b $ (Fig.1). If $ {\cal C} $ is far enough from the hole
the current disappears and we can use Eq.(4) to evaluate the path integral
of $ \nabla \varphi - 2 \pi {\bf A}/\Phi_0 $ along $ {\cal C} $.
Assuming that the
junction has no extension along $ {\bf n} $ we obtain the flux
\begin{equation}
\frac{\Phi}{\Phi_0} = n+ \frac{\chi}{2 \pi} = n + \frac{\chi_0 + \chi_1}{2 \pi}
\end{equation}
\noindent
with $ \chi_0 (= \varphi_b - \varphi_a ) $ determined from the
condition of vanishing current through the junction
using Eq.(3) ($ \chi_0
= {\rm arctan}[(J_{12} - J_{21})/(J_{11} + J_{22})] $) and
\begin{equation}
\chi_1 = \left[\int^{\infty}_{0+} + \int^{0-}_{-\infty} \right] d \tilde{x}
\frac{K'}{R(\tilde{x})} (n^2_x - n^2_y) \sum_{i,j} \varepsilon_{ij}
u_i (\tilde{x}) \frac{\partial u_j }{\partial {\tilde{x}}} (\tilde{x}) .
\end{equation}
\noindent
from Eq.(4).
Since the integrand is finite only near the junction we have restricted the
integration to the straight line part of $ {\cal C} $ on either side of the
junction (Fig.1). It is easy to see that both, $ \chi_0 $ and $ \chi_1 $,
vanish if the system is symmetric with respect to reflection at the
junction ($ \tilde{x} \to - \tilde{x} $) because in this case
$ J_{12} = J_{21} $ and the contribution to the
integrals of Eq.(6) of the two sides cancel each other.
Of course, $ \chi_0 $ and $ \chi_1 $ vanish also if $ {\cal T} $ is conserved
($ \theta = 0 $ and $ \pi $). However, in case that both symmetries are
broken, $ \chi_0 $ and $ \chi_1 $ are
finite and can have any value depending on the system parameters such that
$ \Phi $ is arbitrary and non-zero for any integer winding number $ n $.
The flux $ \Phi' $ in the other hole is given by the requirement that the
sum of both fluxes have to be added to give an integer multiple of $ \Phi_0 $.
Our treatment
applies whenever the superconducting state breaks the time-reversal symmetry
everywhere in the bulk or only locally at the junction.
In either case there would be
a transition temperature $ T^* $, lower than the onset temperature of
superconductivity, above which $ {\cal T} $ is conserved and the flux
quantization of the  holes is standard. Hence, we expect that the flux
$ \Phi $ changes with temperature. In particular, for an experiment
in zero external field $ \Phi $ would be zero for $ T > T^* $ and become
spontaneously finite immediately below $ T^* $.
Our discussion is based on the assumption that the junction is homogeneous
on its entire length.
Because in reality it is difficult to satisfy such a condition,
we eximine briefly
the problem of an inhomogeneous junction where
the coupling $ J_{ij} $ and the order parameter $ u_i(\tilde{x}) $
vary with the position $ \tilde{y} $ along the junction.
Thus, also the two phases $ \chi_0 $ and
$ \chi_1 $ depend on $ \tilde{y} $. The length scale of their
variation has a lower bound at
the coherence length $ \xi $ of the superconductor.
If we assume that $ \chi_0 (\tilde{y}) $ and $ \chi_1(\tilde{y}) $ are fixed,
we may describe the inhomogeneous junction by a the well-known
sine-Gordon equation for the effective
phase difference $ \chi $ of the junction.
\begin{equation}
\frac{\partial^2 \chi}{\partial \tilde{y}^2} =  \lambda^{-2}_J (\tilde{y})
\sin(\chi - \chi_0(\tilde{y}) - \chi_1(\tilde{y}))
\end{equation}
\noindent
where
$ \lambda^{-2}_J= (c \Phi_0/16 \pi \lambda) [ (J_{11}+J_{22})^2 +
(J_{12} - J_{21})^2]^{1/2} $ with $ \lambda $ as the London penetration
depth perpendicular to the junction. The local field on the junction
is given by
\begin{equation}
B_z(\tilde{y}) = \frac{\Phi_0}{4 \pi \lambda} \frac{\partial \chi}{\partial
\tilde{y}}.
\end{equation}
\noindent
Therefore an inhomogeneous junction carries a distribution of
magnetic field whose magnitude depends on the variation of $ \chi $.
The characteristic length scale of variation for $ \chi $ is
$ \lambda_J $ (the screening length along the junction) which is much
larger than $ \xi $. Variations of $ \chi_0 $ and $ \chi_1 $ on the length
scales much shorter than $ \lambda_J $ do not affect $ \chi $. Instead
of following these variation $ \chi $ takes a value which is an
average of $ \chi_0 + \chi_1 $ over a length of the order of $ \lambda_J $.
Therefore
a junction which does not change its properties averaged over a length
$ \lambda_J $ or larger can be considered as homogeneous for our
purpose.
The orientation of the junction within the crystal lattice
has a large influence on the boundary conditions. The
d-wave order parameter is most strongly affected, if $ {\bf n} $ is
close to the (1,1)-direction \cite{12}, so that in case of a
$ {\cal T} $-conserving bulk d-wave supercondutor the conditions to find
local $ {\cal T} $-violation are optimal here \cite{6}. On the other
hand,
we expect that the fluctuations of the junction properties are enhanced
for this direction, because the Josephson coupling for the d-wave component
($ J_{11} $) is reduced. Therefore, it can be characteristic for
junctions with $ {\bf n} $ close to $ (1,1) $ to have a varying
magnetic field distribution over their whole length.
We turn now briefly to the possibility of a device with a variable flux
at fixed temperature.
The arrangement in Fig.1 can be modified by placing a metallic gate over
the whole length of the Josephson junction (Fig.2). A voltage on the
gate changes the tunnelling properties of the junction as well as the
order parameter in its vicinity. Thus the quantity $ \chi_0 $ and $ \chi_1 $
can be tuned by the gate voltage so that a change of $ \Phi $ results
according to Eq.(5). If the bulk superconducting state is
time-reversal symmetric,
the gate may be used to modifiy the junction in order to introduce a
local $ {\cal T} $-violation as described in Ref.5 and 8. In principle this
arrangement
could be used to switch a magnetic field in the holes on and off by
variation of the gate voltage.
In summary, we have discussed an arrangement of holes and a Josephson
junction of mesoscopic length scales, which allows one to identify a
superconductor with broken time-reversal symmetry by observation of
a fractional magnetic flux. Various tools like the SQUID scanning or
the electron holography microscope could be applied for the detection of
such fluxes. Of course, variations of this arrangement are
possible. It may be simplified by expelling one of the two holes out of
the superconductor such that only one hole is remaining which is then
connected by the junction to an edge of the film. The holes may also
be arbitrarily small.
We are grateful to A. Furusaki, J.R. Kirtley,
K. Kuboki, D.K.K. Lee, P.A. Lee, T.M. Rice and K. Ueda for
stimulating
discussions. M.S. acknowledges a fellowship financed by the
Swiss Nationalfonds and Y.B.K. is supported by the NSF grant No.
DMR-9022933.


\vskip 1 cm
\noindent
{\bf Figures Caption}
\vskip 0.6 cm
\noindent
{\bf Figure 1:} Two holes (shaded) connected by a Josephson junction
(dashed line) in a superconducting film. The contour $ {\cal C} $ from
the side a to the side b is used to evaluate the flux in the hole on
the right hand side.
\vskip 0.5 cm
\noindent
{\bf Figure 2:} Schematice cross-section through the junction with a
gate on the top. The asymmetry of the gate ensures that the reflection
symmetry is broken.
\end{document}